# A COMMENT ON TECTONICS AND THE FUTURE OF LIFE ON TERRESTRIAL PLANETS

Milan M. Ćirković

*Astronomical Observatory, Volgina 7,*

*11160 Belgrade, Serbia and Montenegro*

e-mail: arioch@eunet.yu

One of the most important developments in both Earth sciences and astrobiology during the last decade was realization that geophysical processes in general, and plate tectonics and mantle convection in particular, play an important role in the emergence and evolution of life (e.g., Des Marais 1994). This deep interdisciplinary link is splendidly illustrated by the recent discussion of the carbon sequestration limit on the future habitability of Earth (Lindsay and Brasier 2002a,b; Gerstell and Yung 2003). In the original study, Lindsay and Brasier (2002a) suggested that the decrease in geological activity following the depletion of radiogenic isotopes might spell the end of the terrestrial biosphere as a viable system. Gerstell and Yung (2003) argue that this limit is inapplicable to Earth, since the arresting of the tectonic conveyor of carbon will come only after the biosphere is already destroyed by the moist greenhouse effect in about 1.1 Gyr from now. Lindsay and Brasier (2002b) reply that the thrust of their discussion is not the future fate of Earth, but rather the history of planets like Mars which lack the tectonic activity and thus have already become uninhabitable.

Here we would like to take a look at this issue from a wider perspective in both spatial and temporal sense. The crucial lead comes from the emphasis of Lindsay and Brasier on the possible histories of life in other places, such as Mars. The issue of the future evolution of the terrestrial biosphere—and other biospheres, as we shall discuss below—is an important part of the nascent discipline of physical eschatology, dealing with future evolution of celestial bodies (e.g., Adams and Laughlin 1997; Ćirković 2003). A natural generalization of the terrestrial biosphere is the **Galactic habitable zone** (GHZ; cf. Gonzalez, Bronwlee, and Ward 2001) which comprises a vast annular ring wherein all inhabitable planets in the Milky Way galaxy are located. If we reject simplistic geocentrism, we are now in good position to assess prospects for existence of viable biospheres elsewhere in the Milky Way. In this note we would like to emphasize the truly global importance of geological activity all over GHZ. As Ward and Brownlee (2000) hint—but not fully explicate—in their celebrated monograph *Rare Earth*, GHZ also has a finite lifetime. Interestingly enough, the main reason for this finite lifetime may well be geological and not (primarily) astrophysical: the arrest of geological activity all over the Galaxy. Within such a perspective we can agree with both Lindsay and Brasier that the radiogenic heat imposes a limit on the long-term survival of planetary biospheres **and** Gerstell and Yung that such limit is inapplicable in the specific case of Earth. However, this does not mean that the limit is unimportant; it can indeed determine life and death on millions of other worlds.

It is rather easy to conclude that geological activity of terrestrial planets has, in fact, been stronger in the Galactic past. This is due to the increased chemical abundances of radioactive elements, notably U, Th, and isotope $^{40}$K, main generators of planetary geological activity and its interaction with the atmosphere through the carbon cycle. It has been shown recently that most terrestrial planets in the Milky Way Galaxy are significantly older than Earth; in fact, the average age of a terrestrial planet is 6.4 ± 0.9 Gyr, according to the metallicity calculations of Lineweaver (2001). Radiogenic isotopes generating and sustaining geological activity in terrestrial planets are *r*-process elements created predominantly by neutron capture in Type II supernovae, signaling death of very massive and luminous stars (e.g. Clayton 1983). Although the exact rate of supernovae in our Galaxy is still subject to some controversy (due mainly to the obscuration by dust of most of the Milky Way stellar disk), there is incontrovertible astrophysical evidence that the supernovae rates in spiral galaxies in general decrease steadily with cosmic time (Yungelson and Livio 2000). The reason for this is rather simple: general star formation rate decreases with cosmic time due to the consumption of interstellar gas and only very imperfect recycling of it in this "galactic ecology".

Now, if we accept the assumption of Ward and Brownlee, as well as Lindsay and Brasier, that geological activity in general, and plate tectonics in particular are **essential** for the development of complex metazoan life on **any** terrestrial planet, pessimistic conclusions are inescapable. Very early in the history of the Galaxy, terrestrial planets could not form at all, due to the low metallicity of protoplanetary material. Very lately in the history of the Galaxy, any terrestrial planet formed will lack radiogenic isotopes, and is bound to become a geologically dead body at timescales of ~1 Gyr after its formation. Thus, subsequent planetary evolution will lack major mechanisms for carbon burial and oxygenation, which will, in turn, adversely influence any prospects of life evolving complex metazoan forms. If the case of Earth is typical with its large difference between the epoch of biogenesis and the development of such complex lifeforms ("Cambrian explosion")—the interval of more than 3 Gyr—it seems clear that the future planets poor in radioactive elements will have negligible chances of replicating the levels of complexity and diversity present on Earth. Therefore, it seems that this "window of opportunity" is rapidly closing with advance of the nucleocosmochronological clock. Unfortunately, we still cannot pinpoint the exact future epoch at which this tectonic limit will be reached for the Milky Way GHZ due to essentially the same uncertainties present in the terrestrial case: it is unclear which is the minimal amount of radiogenic isotopes sufficient for sustaining the plate tectonics, as well as how important effects of alleged local protosolar metallicity enhancement or chemical fractionation in protoplanetary matter are. It is our hope that future studies will reduce these uncertainties and enable the exact determination of the life span of GHZ.

In any case, we have an elegant and instructive sequence of causal influences: *fundamental physics* (properties of radionuclides) → *astrophysics* (SNe nucleosynthesis) → *astrochemistry and celestial mechanics* (terrestrial planet formation) → *geology* (mantle convection, plate tectonics) → *atmospheric sciences* (greenhouse gases) → *biology* (maintenance of the biosphere). This is an interesting twist, instructive in both heuristical and pedagogical senses, illuminating the tight interconnections between the various levels of the complex system we call the universe.

Why is the tectonic limit inapplicable to the Earth, but applicable to the Galaxy? The explanation lies in an observation selection effect (comprising a part of the Weak Anthropic Principle). Rapid biogenesis on Earth (≥ 3.8 Gyr BP) and subsequent complexification (~ 600 Myr BP)—rapid in comparison to the pace of chemical evolution of the Galaxy—indicate that the Earth will fall within a subset of planets retaining sufficient geological activity to reach the stage of intelligent observers. Once that stage is reached, those observers (us) ought not be surprised to find themselves on a geologically active planetary body, since that is the only state-of-affairs compatible with their own existence. The correct path is to use the Bayesian formalism to attempt to estimate how really typical our situation is in a wider distribution of initial conditions for planetary formation (cf. Bostrom 2002). With tremendous advances of both astrophysics and geochemistry, this will hopefully soon be within our grasp.

Finally, the statement of Gerstell and Yung (as well as similar sentiments expressed in Ward and Bronwlee [2002]) about our fast death following the breakdown of the carbon cycle seem premature and unnecessarily pessimistic. Many technologies whose timescales of development are many orders of magnitude smaller than either astrophysical or geological timescales will enable our descendants efficient reversal of adversary trends (both natural and anthropogenic). Already envisaged macro-projects of geoengineering (e.g. Govindasamy and Caldeira 2000) could easily provide the solution for both anthropogenic global warming and longer-term cooling tendencies related to the end of current interglacial period. By the same token, even longer-term adversary changes such as arresting the plate tectonics and the associated carbon sequestration could be dealt with in a similar manner. The ratio of timescales for geophysical and technological changes experienced so far is so huge that such intentional extension of our capacities to influence the physical environment—envisaged long ago by such visionaries like Percy B. Shelley and Herbert G. Wells—cannot be excluded in any way.

**References**

Adams, F. C. and Laughlin, G. 1997, A dying universe: the long-term fate and evolution of astrophysical objects. *Rev. Mod. Phys.* **69**, 337-372.

Bostrom, N. 2002, *Anthropic Bias: Observation Selection Effects in Science and Philosophy* (Routledge, New York).

Ćirković. M. M. 2003, Resource Letter PEs-1: Physical eschatology. *Am. J. Phys.* **71**, 122-133.

Clayton, D. D. 1983, *Principles of Stellar Evolution and Nucleosynthesis* (University of Chicago Press, Chicago).

Des Marais, D. J. 1994, Tectonic control of the organic carbon reservoir during the Precambrian. *Chem. Geol.* **114**, 303-314.

Gerstell, M. F. and Yung, Y. L. 2003, A comment on tectonics and the future of terrestrial life. *Precambrian Res.* **120**, 177-178.

Gonzalez, G., Brownlee, D., and Ward, P. 2001, The Galactic Habitable Zone: Galactic Chemical Evolution. *Icarus* **152**, 185-200.

Govindasamy, B. and Caldeira, K. 2000, Geoengineering Earth's radiation balance to mitigate $CO_2$-induced climate change. *Geophys. Res. Lett.* **27**, 2141-2144.

Lindsay, J. F., and Brasier, M. D. 2002a, Did global tectonics drive early biosphere evolution? Carbon isotope record from 2.6 to 1.9 Ga carbonates of Western Australia basins. *Precambrian Res.* **114**, 1-34.

Lindsay, J. F., and Brasier, M. D. 2002b, A comment on tectonics and the future of terrestrial life—reply. *Precambrian Res.* **118**, 293-295.

Lineweaver, C. H. 2001, An Estimate of the Age Distribution of Terrestrial Planets in the Universe: Quantifying Metallicity as a Selection Effect. *Icarus* **151**, 307-313.

Ward, P. D. and Brownlee, D. 2000, *Rare Earth: Why Complex Life Is Uncommon in the Universe* (Springer, New York).

Ward, P. D. and Brownlee, D. 2002, *The Life and Death of Planet Earth: How the New Science of Astrobiology Charts the Ultimate Fate of Our World* (Henry Holt and Company, New York).

Yungelson, L. R. and Livio, M. 2000, Supernova rates: A cosmic history. *Astrophys. J.* **528**, 108-117.